# Leadership and Participation in NASA's Astrophysics Explorer-Class Missions

*Astro2020 State of the Profession Considerations White Paper*

Joan Centrella[1,2], Michael New[3], and Meagan Thompson[3]

**Abstract:**

We have conducted a data study of leadership and participation in NASA's Astrophysics Explorer-class missions for the nine solicitations issued during the period 2008-2016, using gender as a marker of diversity. During this time, 102 Principal Investigators (PIs) submitted Explorer-class proposals; only four of these PIs were female. Among the 102 PIs, there were 61 unique PIs overall; of these, just three were female. The percentage of females in science teams in these proposals ranges from a low of 10% to a high of 19% across the various solicitations. Combining data from all these Explorer-class proposals, we find that the overall participation by females in science teams is 14%. Eighteen of the Explorer-class proposals had zero females in science roles, and this includes science teams with as many as 28 members. These results demonstrate that participation by women in the leadership of and, in many cases, on the science teams of proposals for Explorer-class missions is well below the representation of women in astronomy and astrophysics as a whole. In this white paper, we present our data and a discussion of our results, their context, and the ramifications for consideration by Astro2020 in its study of the state of the profession.

1 Astrophysics Division, Science Mission Directorate, NASA Headquarters
2 Astrophysics Science Division, NASA's Goddard Space Flight Center; contact email: joan.centrella@nasa.gov
3 Science Mission Directorate, NASA Headquarters

**Key Issue and Overview of its Impact on Astronomy and Astrophysics**

NASA's Astrophysics Explorer-class missions are led by Principal Investigators (PIs), powered by innovation, and designed to enable quick responses to developing science areas between decadal surveys [1, 2, 3]. However, participation by women in the leadership and, in many cases, on the science teams of proposals for Explorer-class missions is well below the representation by women in astronomy and astrophysics as a whole [1, 4, 5]. The data study presented here illustrates this stark lack of diversity. During the period 2008–2016, 102 proposals were submitted for Explorer-class missions; only four of these proposals had a female PI (**Figure 1**). Of the full set of 102 proposals, there were 61 unique PIs; of these, only three were female. Each proposal has a science team, including Co-Investigators (Co-Is) and Collaborators across the range of science needed for mission success. Eighteen of the proposals had zero females in science roles (**Figure 4**), for science teams with as many as 28 members (**Figure 5**).

This data reveals a significant lack of participation by females in key NASA astrophysics missions. *Why does this occur, and to what extent is this a problem?* Lack of gender diversity can signal the absence of other types of diversity as well [6, 7, 8]. Research shows that team excellence and diversity go together, especially for innovative activities [1, 9, 10]. In addition, inclusive behaviors and practices—not just diversity—are essential to create a sense of community and enable all team members to contribute their best [1, 11]; these are key ingredients for mission success.

The Statement of Task [12] for the Astronomy and Astrophysics Decadal Survey (Astro2020) directs the survey to:

> *Assess the state of the profession, using information available externally and, if necessary, data gathered by the study itself, including workforce and demographic issues in the field. Identify areas of concern and importance to the community raised by this assessment in service of the future vitality and*

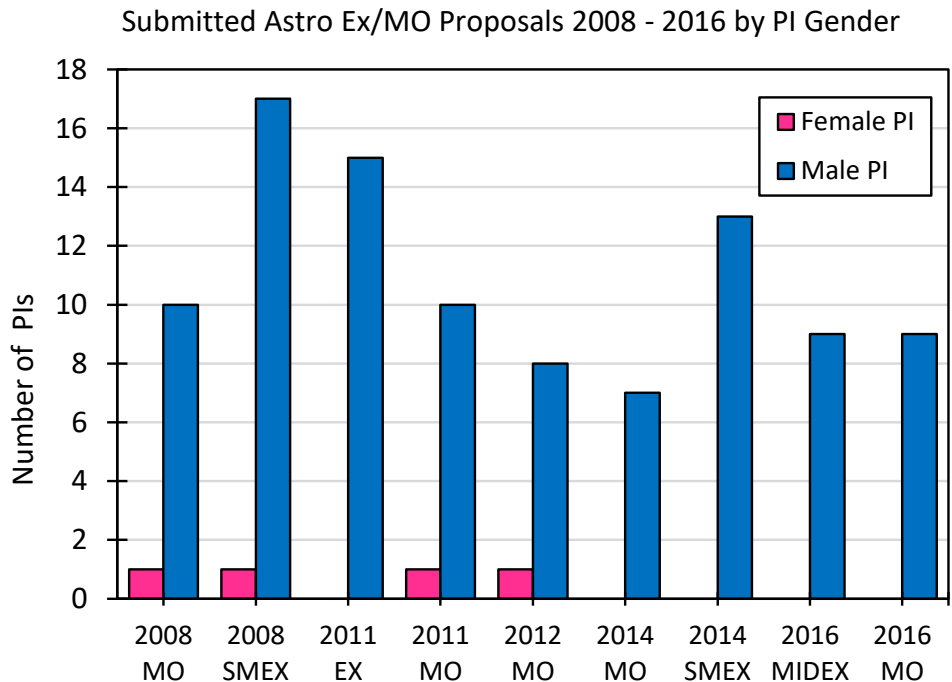

**Figure 1.** *Distribution of PIs for proposals submitted to Astrophysics Ex/MO AOs during 2008–2016 by number and gender.*

*capability of the astronomy and astrophysics work force. Where possible, provide specific, actionable and practical recommendations to the agencies and community to address these areas.*

In this white paper, we present our data; describe our methods; and discuss our results, their context, and the ramifications for consideration by Astro2020 in its study of the state of the profession.

**Scope of this Study**

We focus on proposals submitted to Announcements of Opportunity (AOs) for NASA's Astrophysics Explorer Program during the period 2008–2016. There are two basic categories of Explorer-class missions; see, e.g., [13] and links therein. Stand-alone missions, with cost caps ranging from ~ $100M to ~ $250M include the Small Explorer (SMEX), Explorer (EX) and Medium Explorer (MIDEX) categories. Smaller investigations, called Missions of Opportunity (MOs), have cost caps ranging from ~ $35M for suborbital-class missions (including CubeSats and SmallSats) to ~ $70M for payloads attached to platforms such as the International Space Station or non-NASA satellites. These AOs are summarized in **Table S-1** in the supplementary material [14].

To examine career paths to leadership and participation in these Ex/MO proposals, we looked at proposals submitted to NASA's Astrophysics Research Opportunities in Space and Earth Science (ROSES) elements. Given the importance of technological innovation for Ex/MO proposals, we focused primarily on the Astrophysics Research and Analysis (APRA) Program during 2006–2017 and the Strategic Astrophysics Technology (SAT; this includes the Technology Development for Exoplanet Missions (TDEM)) Program during 2010–2017. We also looked briefly at the Astrophysics Data Analysis Program (ADAP), the Astrophysics Theory Program (ATP), and the Nancy Grace Roman Technology Fellowships (RTF) Program.

We obtained our data from three sources. The NASA Solicitation and Proposal Integrated Review and Evaluation System (NSPIRES) internal user reporting tools provided solicitation (AO) and proposal identifiers as well as full names, team member roles and institutions for all proposal personnel. We used gender as a measure of diversity, because it is the factor we can most readily infer[1]. Gender data was obtained using the online tool Gender API [15], which uses first names to infer binary female/male (F/M) gender from a database of over 2.1 million names, covering 178 countries. Team members having first (given) names with reported accuracy of 95% or higher for inferred gender were assigned to the binary F/M categories; names with lower accuracy ratings were assigned to the categories using public domain searches. Finally, to indicate the career stage for team members, we used public domain searches of curriculum vitae (CVs) to determine the number of years since PhD for each proposal.

**Analysis and Results**

Our data comprises a total of 102 Ex/MO proposals over nine AOs. **Figure 1** shows the distribution of PIs by inferred gender (hereafter, gender) for each of these AOs; this information is shown in tabular form in **Table S-2** in the supplementary material [14]. Overall there are 102 PIs, and 61 unique PIs, with 25 PIs (one female, 24 males) serving as an Ex/MO PI more than once. Females comprise 4% of the PIs overall, and 5% of the unique PIs; see **Table S-3** in [14].

The distribution of PhD ages for the Ex/MO PIs is shown in **Figure 2**. The median PhD age for all PIs at submission is 25.5 years. For female PIs the median PhD age at submission is 18.5 years; since the sample of female PIs is very small, this difference may not be significant.

---
[1] We use inferred gender because we do not have user-supplied data for our sample. Future studies will incorporate user-supplied information for a broader range of diversity factors.

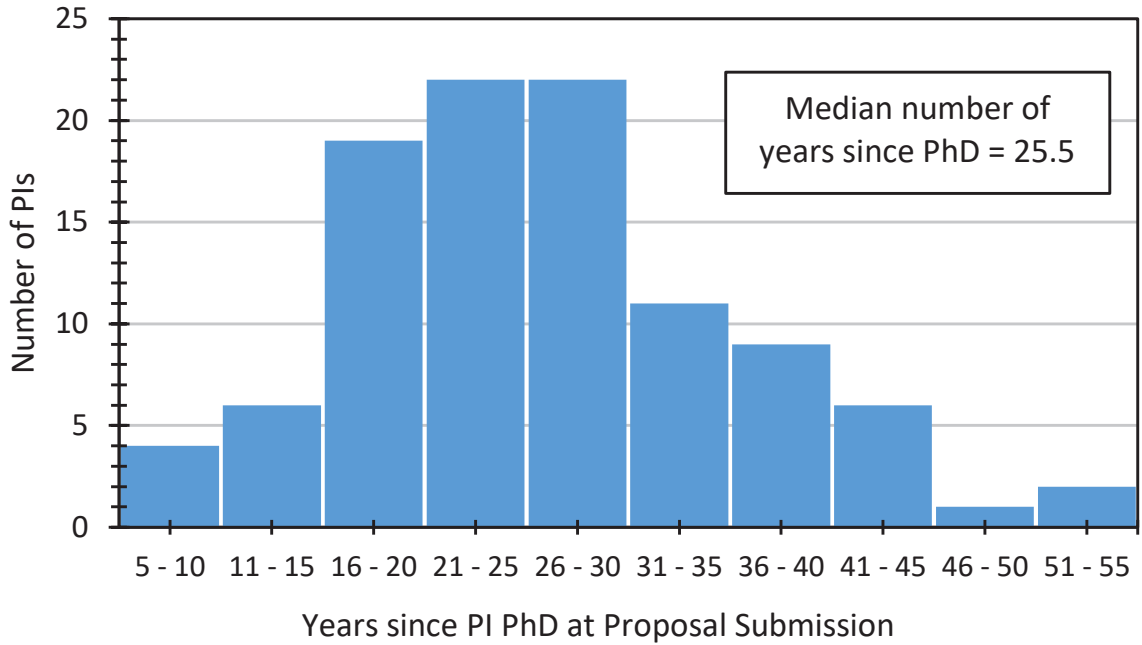

**Figure 2.** *Distribution of PhD ages for PIs of all Ex/MO proposals during 2008–2016. The PhD age is calculated at the year of proposal submission.*

**Figure 3** shows the overall distribution of submitting organization types. For our purposes, the Jet Propulsion Laboratory (JPL) is included in the same category as NASA Centers. The designation "Other Federal Agencies" includes government labs and federally funded research and development centers (FFRDCs) other than JPL. Universities submitted 51, or 50%, of the proposals overall. More details are given in **Figure S-1** in [14], which shows the breakdown of submitting institutions by opportunity and organization type. Most of the PIs, and all of the female PIs, remained at and submitted their proposals from the same institution throughout 2008–2016. However, three PIs changed both institutions and institution types, and one changed institutions within the same type. Additionally two PIs, while each remaining at his home institution (a university), changed his submitting institution name and type: one submitted his first proposal from his home university and later proposals from a NASA center, while the other submitted his first proposal from a NASA center and later ones from his home university.

We look next at the composition of Ex/MO proposal science teams, with the focus on participation by gender. We define the *science team* to comprise these 11 team member roles: Co-I, Co-I/Co-PI (non-US organization only), Co-I/Institutional PI, Co-I/Science PI, Collaborator, Deputy PI, Instrument Scientist, International Partner, PI, Postdoctoral Associate, and Project Scientist. A Co-I is a team member who plays a necessary role in the proposed investigation (whether funded by the Astrophysics Explorers Program, or contributed), whereas a collaborator is an individual who is less critical to the successful development of the mission than a Co-I and cannot be funded by the Explorer Program [13]. Note that the AOs only define the PI, Co-I, Collaborator, and Project Scientist roles [13]; the other categories of member roles are defined within the text of individual proposals, and are not necessarily the same in all cases[2]. In addition, our data includes only the number of individuals in the named roles, and therefore does not account for the percentage effort relative to a full-time equivalent (FTE) each member committed to the proposed work.

---

2 In particular, the Co-I/Science PI role for an Ex/MO proposal does not have the same meaning as in ROSES proposals (in which this role is held by a person who is in charge of the science investigation but unable to serve as PI for institutional reasons, and an institutional representative then serves as the PI); see the NASA Guidebook for Proposers [16].

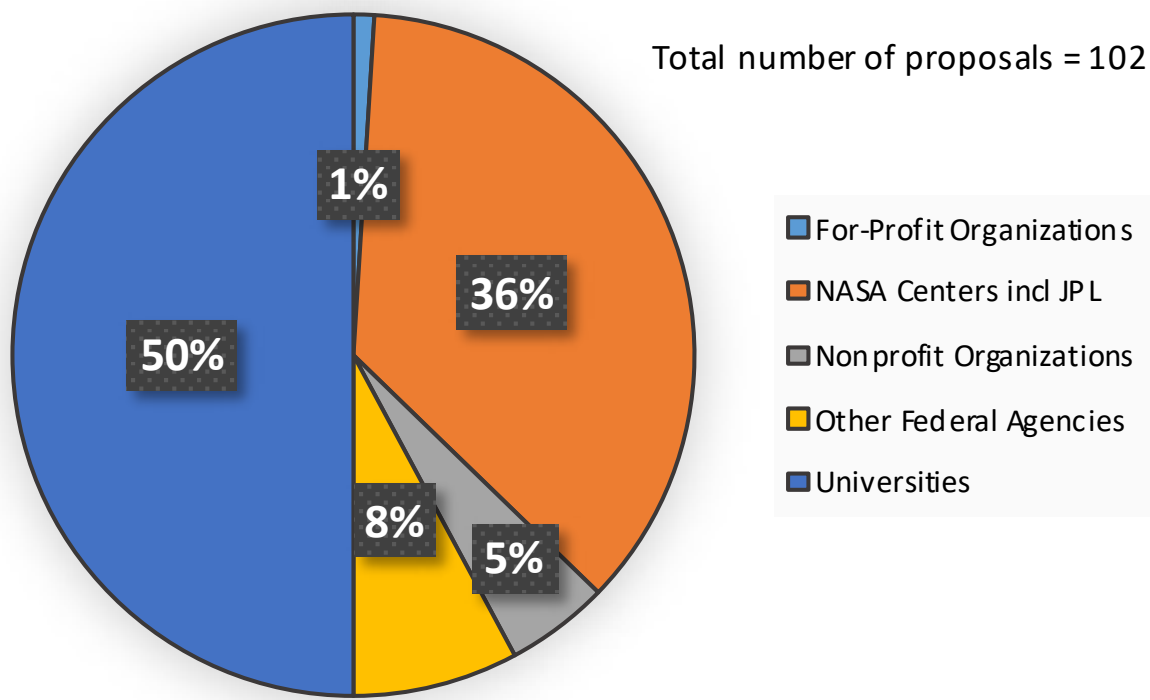

**Figure 3.** *The overall distribution of submitting organization types for Astrophysics Ex/MO proposals 2008–2016.*

The size of the science teams varies widely, from a minimum of six to a maximum of 77 members; see **Table S-4** in [**14**]. For all 102 proposals, the median size of science teams is 18 members. Examining the categories of opportunities individually, we did not find any statistically significant difference in the size of science teams when comparing among MOs, SMEXs, and MIDEXs (including EXs).

**Table 1** shows the distribution of team members among science roles for the entire group of 102 proposals broken down by gender. The overall distribution of team members in the science roles is displayed as a pie chart in **Figure S-2** in [**14**]. All proposals have one PI, at least several Co-Is, and one or more Collaborators. The other science roles are not assigned in all of the proposals. The 2014 SMEX and 2016 MIDEX AOs recommended (but did not require) the designation of a Deputy PI [**13**]. This role was used only in four proposals submitted to the 2016 MIDEX and 3 proposals submitted to the 2016 MO.

For all Ex/MO proposals taken together, the overall participation in science roles is 14% for females and 86% for males. Considering each of the AOs separately, we see that overall female participation ranges from a low of 10% for the 2011, 2012, and 2014 MOs to a high of 19% for the 2016 MIDEX and 2016 MO; see **Figure S-3** in [**14**].

The percentage of science participation by females varies even more widely among the set of individual proposals. In fact, *18% of the proposals have zero females in science roles*. This is shown in **Figure 4**, where the number of proposals is plotted against the number of females in science roles. *Proposals with zero females in science roles constitute the peak of the distribution*, and the distribution decreases monotonically as the number of females in science roles increases. For the set of 102 proposals taken together, the median number of females in science roles is 3.5. Males-only proposals were submitted in every solicitation from 2008–2016 except the for 2016 MIDEX; see **Figure S-4** in [14]. Three organizational types submitted males-only proposals: universities submitted 67% of these, followed by NASA Centers including JPL (28%), and Other Federal Agencies (5%); see **Figure S-5** in [14].

**Table 1.** *Science team roles by gender for all Astrophysics Ex/MO proposals 2008–2016 combined.*

| Science Role | # F | # M | Total # |
|---|---|---|---|
| Co-I | 195 | 1101 | 1296 |
| Co-I/Co-PI (non-US organization only) | 2 | 18 | 20 |
| Co-I/Institutional PI | 5 | 92 | 97 |
| Co-I/Science PI | 2 | 1 | 3 |
| Collaborator | 83 | 459 | 542 |
| Deputy PI | 2 | 5 | 7 |
| Instrument Scientist | 1 | 11 | 12 |
| International Partner | 0 | 10 | 10 |
| PI | 4 | 98 | 102 |
| Postdoctoral Associate | 2 | 9 | 11 |
| Project Scientist | 2 | 18 | 20 |
| **Grand Total** | **298** | **1822** | **2120** |

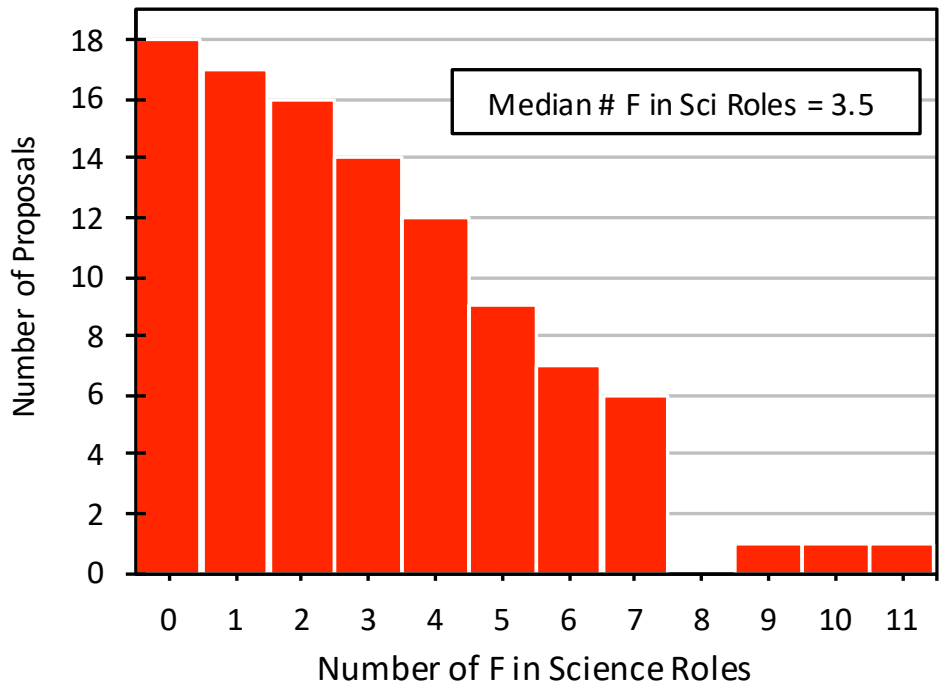

**Figure 4.** *The number of proposals versus the number of females in science roles is shown. Note that 18% of the submitted Ex/MO proposals have zero females in science roles; this is the peak of the distribution.*

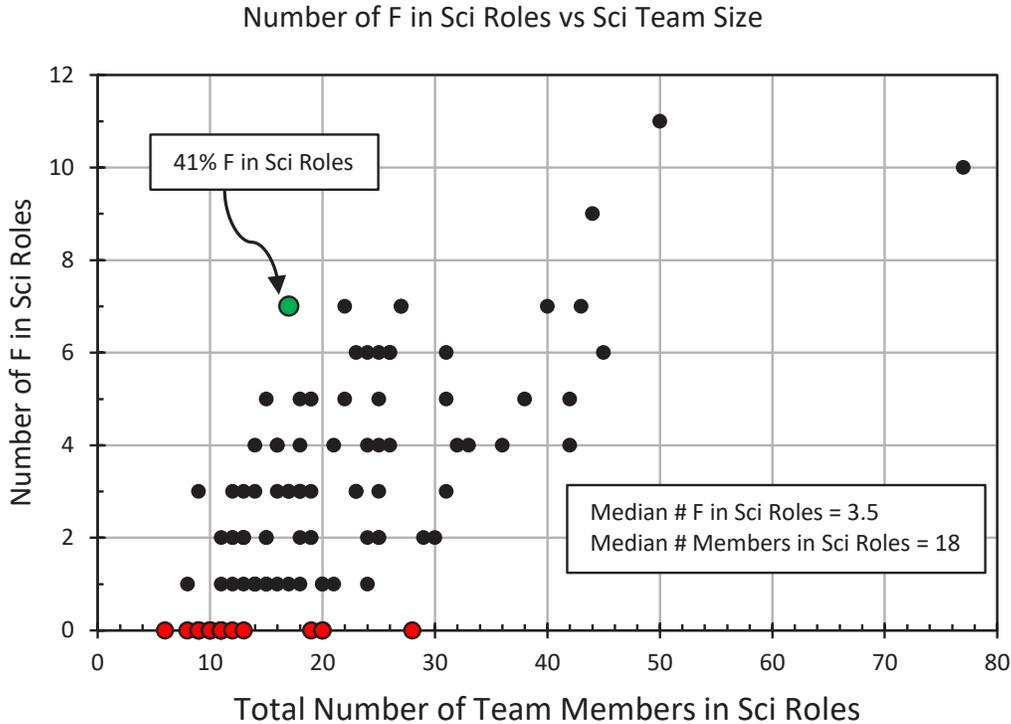

**Figure 5.** *The number of females in science roles is shown against the size of the science team for all 102 Ex/MO proposals submitted during 2008–2016. Note the presence science teams with as many as 28 members that have zero females (red dots). The proposal with the highest percentage of females, 41%, is shown by the large green dot; this was submitted by a PI at a university to the 2008 SMEX AO.*

The number of females in science roles on each Ex/MO proposal is plotted against total science team size in **Figure 5**. There is some correlation between these two variables, with the number of females in science roles trending generally higher as the size of the science team increases (correlation coefficient ~ 0.7). We see that *during 2008–2016, there are Ex/MO science teams with as many as 19, 20 and even 28 members that have only males on their science teams*.

Next, we examine ROSES proposals with the focus on technology proposals submitted to APRA and SAT. **Figure 6** shows the overall participation in APRA/SAT proposals during 2006–2017 for all 61 unique Ex/MO PIs. Here, all science team roles held by the Ex/MO PIs are included in the total of 833 APRA/SAT proposal roles. Note the relative importance of the Suborbital Program shown by the blue wedge; this includes balloons and sounding rockets. The purple Suborbital CubeSat wedge is much smaller, since this element only began in 2013. The grey wedge labeled "Other Proposal Types" includes proposals in the Laboratory Astrophysics and Ground-based Observations APRA categories, as well as those without a specified classification. Key science proposal roles are broken out in **Figure S-6** in [14], which shows the large number of PI and Co-I roles in APRA Suborbital proposals held by Ex/MO PIs.

The APRA/SAT participation for the three unique female PIs looks rather different. The number of roles per Ex/MO female PI ranges from zero to 13. For this group, there were 17 APRA/SAT roles; of these, eight were in the APRA Supporting Technology category and three were in Suborbital. We caution that this group is very small, and the numbers reported here may not be representative of the overall group of female APRA/SAT PIs who may have considered proposing to an Ex/MO AO but, for whatever reasons, did not.

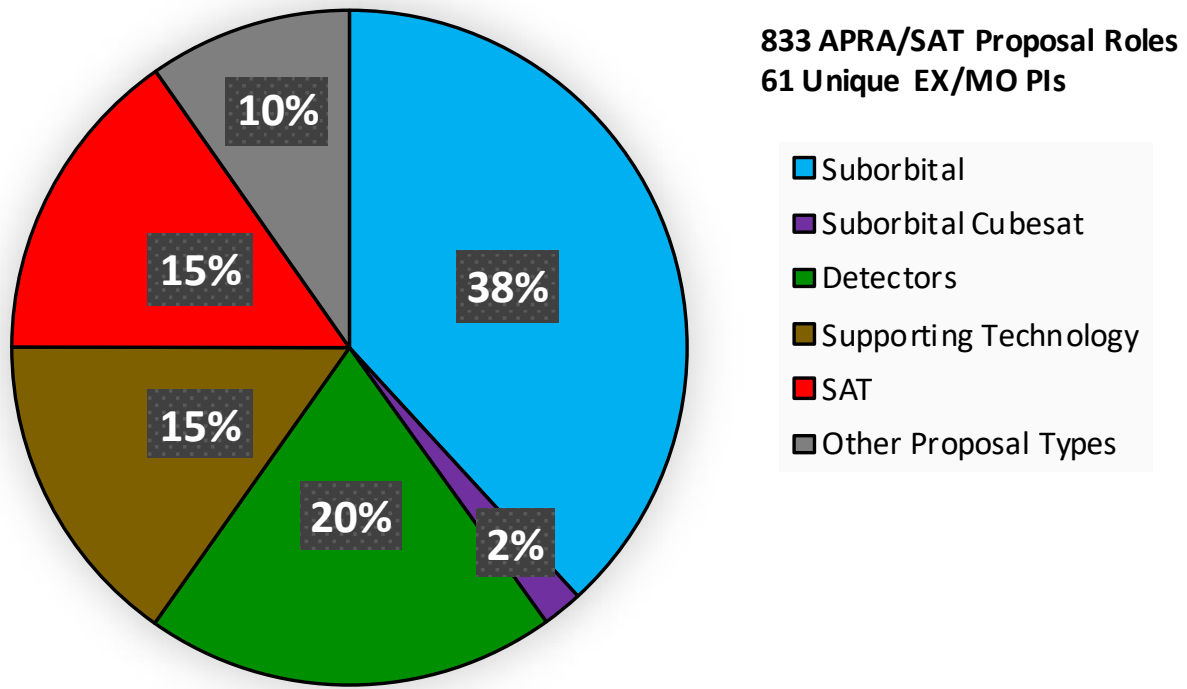

**Figure 6.** *Overall participation in APRA/SAT during 2006–2017 for the 61 unique Ex/MO PIs. All APRA/SAT science team proposal roles are included.*

Leadership and participation in the APRA/SAT Programs are important, but not definitive, factors for the 61 unique Ex/MO PIs. We find that:

- Only eight (8%) of the Ex/MO PIs have zero APRA/SAT science roles;
- The median number of APRA/SAT science roles held by an Ex/MO PI is 22;
- The largest number of APRA/SAT roles is 43 (held by two Ex/MO PIs);
- Thirty-seven (61%) were PIs on at least one APRA/SAT proposal during 2006–2017;
- The remaining 24 (39%) did not submit an APRA/SAT proposal during this period; however, they may have held other science team roles on APRA/SAT proposals;
- Nine (38%) of these 24 served as PI on more than one Ex/MO proposal;
- For PIs on Ex/MO proposals submitted from 2011 to 2017, 55% of them received APRA/SAT awards prior to being a mission PI (i.e., during 2008–2010);
- Fifty-one (84%) of the 61 unique Ex/MO PIs were members of the science team on one or more APRA/SAT proposal during 2006–2017.

We also compare the participation by gender of the Ex/MO PIs and science team members with a broader segment of the Astrophysics ROSES programs. **Figure 7** shows that the percentage of female PIs submitting APRA, ADAP, and ATP proposals for the time periods studied (17%) is somewhat larger than but still comparable with the overall percentage of females in Ex/MO science roles (14%), reflecting the breadth of expertise expected on Ex/MO science teams. While the percentage of female APRA PIs (9%) is lower than in both these cases, it is more than twice the percentage of female Ex/

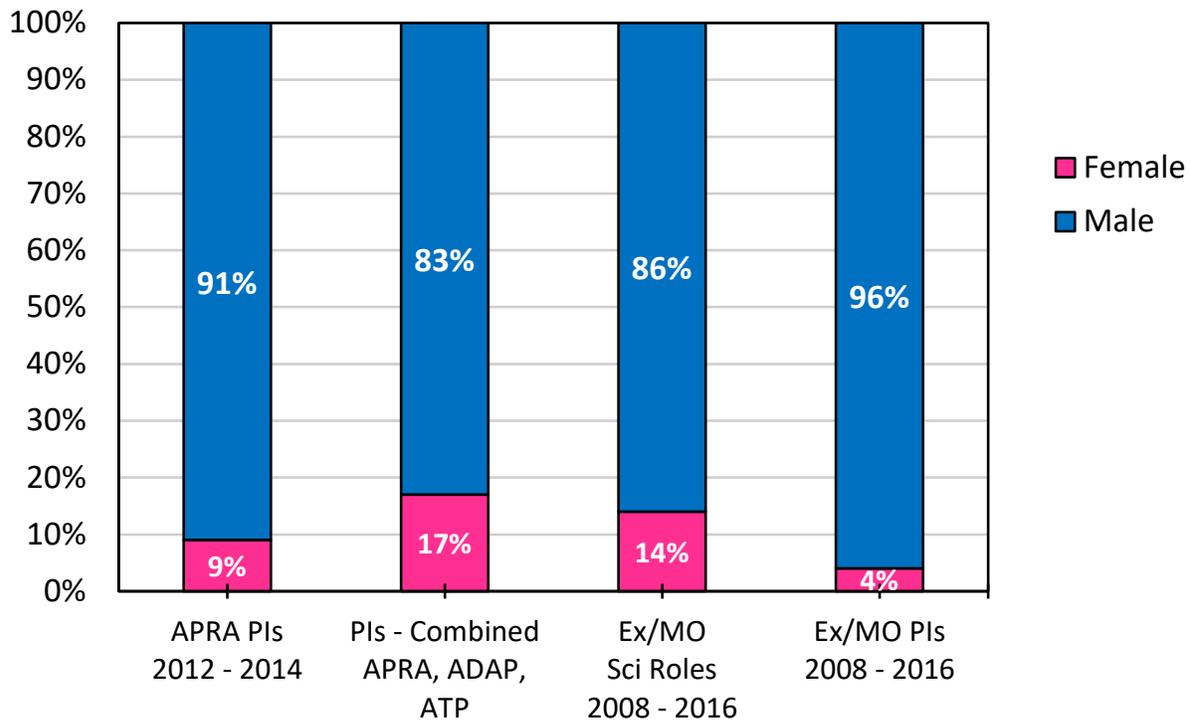

**Figure 7.** *Comparison of ROSES PIs with EX/MO PIs and EX/MO science team members by gender. The column shown for each category reaches to the 100% level to encompass all proposals submitted. Within each bar, the percentage of PIs by gender is shown in the color-coded segments relative to the y-axis, and the percentage of PIs in each case is printed within the segments. The ROSES data was compiled by Dan Evans. The ATP and APRA data is from 2012–2014; the ADAP data is from 2013–2015. Note that the APRA data does not include SAT.*

MO PIs (4%). The ROSES data is shown in more detail in **Table S-5** in [**14**]. Note that while the ADAP and ATP Programs show similar percentages for submitted and selected proposal with a female PI, the APRA Program shows a bigger drop from submitted (9%) to selected (5%) proposals.

Looking toward the future, the goals of the RTF Program are to give early career researchers the opportunity to develop the skills necessary to lead astrophysics flight instrumentation development projects and become PIs of future astrophysics missions; to develop innovative technologies that have the potential to enable major scientific breakthroughs; and to foster new talent by putting early-career instrument builders on a trajectory towards long-term positions [**17**]. This program began in 2011, and was restructured in 2017 to work in conjunction with the APRA Program [**18**]. The data on the number of PIs by gender for initial proposal submission, intermediate selection, and final selection as a Roman Technology Fellow is shown in **Table S-6** (2011–2015) and **Table S-7** in [**14**]. Combining all years, 62 initial proposals were submitted to the RTF Program, of which 13 (21%) had a female PI. At the final step, a total of 12 Roman Technology Fellows were selected, two (17%) of whom are female.

**Discussion**

Recent demographic studies of astronomy and astrophysics reveal that the percentage of females in these fields is significantly larger than what we find for Ex/MO PIs. A 2016 study of AAS members reports that 26% of AAS members overall are female; see **Table 20** and **Figure 12** in Ref. [**5**]. The 2019 Survey of Women in Physics and Astronomy by the American Institute of Physics (AIP) reports the percentage of Physics department faculty members who are females is 16% overall; see **Table 4** in Ref. [4]. In particular, females are 18% of Associate Professors and 10% of Full Professors; these are the cohorts most expected to serve as Ex/MO PIs, based on the distribution of PhD ages in our **Figure 2**. For Astronomy departments, they find 19% of faculty members overall are female, with 15% of full professors and 29% of associate professors; see **Table 5** in [4]. In all cases, these numbers are much higher than the 4% of Ex/MO PIs in our data, although they are closer to the 14% of females participating in Ex/MO science team roles overall. This is consistent with the broader range of backgrounds and expertise represented on the Ex/MO science teams as compared to the stronger emphasis on technology roles—and hence participation in APRA—for the Ex/MO PIs. However, the question still remains: *Why are there so few females in technology areas, as measured by the very low percentages of female Ex/MO and APRA PIs?*

The Assistant Professor ranks in Physics departments (23% female) and Astronomy Departments (29%) have higher percentages of females than the more senior ranks [4]; this is also seen in Ref. [5], which shows that 46% of AAS members born in 1983 or after are female. We can compare these numbers with the early-career RTF PIs (21% female) and awardees (17% female). We caution that the RTF numbers are very small; nevertheless, the presence of some awards to females in this prestigious program may encourage others to consider careers in technology areas.

Our most disturbing finding is the relatively large percentage of Ex/MO science teams that have no females in science roles at all. These 18 males-only proposals comprise the largest subgroup of Ex/MO proposals when ranked by the number of females on the science teams; they include teams with as many as 19, 20, and even 28 members. Males-only proposals were submitted to all AOs from 2008 to 2016, except for the 2016 MIDEX AO. These are all recent proposals, submitted in the early part of the 21st century. They come mainly from universities (67% of all males-only proposals, and 24% of all university proposals), followed by NASA Centers including JPL (28% of males-only proposals, and 14% of all NASA Center proposals). We see from the broader ROSES data that there are substantial numbers of females in programs with expertise relevant to various components of Ex/MO science teams, such as ADAP and ATP. The AAS and AIP data also reveal that there are significant and growing numbers of women in astronomy and astrophysics overall. *Why are so many Ex/MO science teams exclusively males-only, and what are the consequences for the field?*

We note that data about certain important aspects of the institutional context underlying Ex/MO proposals is not available to our study. Ex/MO proposals typically have long lead-times, and are costly for institutions to develop and submit [1]. Only certain institutions, with the requisite laboratories, technical staff, and management support can realistically propose [1]. Such institutions typically invest in current and future PIs through internal research and development (IRAD) funds to mature concepts and technology from APRA/SAT to Ex/MO levels, and with bid and proposal (B&P) funds to develop and submit Ex/MO proposals. Institutional priorities and competitive postures drive decisions about where to make these investments, and which PIs or potential PIs to invest in. And the institutional climate plays a large role in determining which people consider proposing, and which of these can rise to the level of an Ex/MO PI.

Several factors bear consideration. In particular,

- Research shows that team excellence and diversity go together, particularly in innovative activities [1, 9, 10]. Success in Ex/MO proposals depends crucially on team leadership and cohesion [1, 3]. Excellent teams require diverse perspectives and views to avoid group-think. In addition, inclusive behaviors and practices are essential to create a sense of community and enable all team members to contribute their best [1, 11].

- Moreover, the lack of gender diversity can signal the lack of other types of diversity [6, 7, 8]. In particular, recent work demonstrates that women of color face greater risks of gendered and racial harassment [19, 20].
- The recent National Academies study [6] finds that sexual harassment is widespread in academic science and undermines women's professional and educational attainment. It is most likely to take place in environments that are male-dominated in number, leaders, and culture. Organizational climate is, by far, the greatest predictor of the occurrence of sexual harassment.

Our work is part of a larger ongoing exploration of career pathways being carried out by M. New and M. Thompson, Science Mission Directorate, NASA Headquarters. In the study presented here, we focus on leadership and participation in NASA's Astrophysics Explorer-class missions, using gender as a lens with which to view diversity. Our analysis reveals some disturbing findings and raises a number of important questions. We present this white paper to Astro2020 as a window into aspects of the broader astronomy and astrophysics landscape, for consideration in their study of the State of the Profession.

# Supplementary Material for
# *Leadership and Participation in NASA's Astrophysics Explorer-Class Missions*

## *Astro2020 State of the Profession Considerations White Paper*

Joan Centrella[1,2], Michael New[3], and Meagan Thompson[3]

This is supplementary material in support of the Astro2020 State of the Profession Considerations White Paper *Leadership and Participation in NASA's Astrophysics Explorer-Class Missions.*.

___________
1 Astrophysics Division, Science Mission Directorate, NASA Headquarters
2 Astrophysics Science Division, NASA's Goddard Space Flight Center; contact email: joan.centrella@nasa.gov
3 Science Mission Directorate, NASA Headquarters

**Table S-1.** *Astrophysics Explorer-class Missions 2008–2016*

| AO Category | Year | PI-Managed Mission Cost Cap |
|---|---|---|
| SMEX | 2008 | $105M, not including Expendable Launch Vehicle (ELV) (FY08 $) |
| MO | 2008 | $70M (FY08 $) |
| EX | 2011 | $200M, not including ELV (FY11 $) |
| MO | 2011 | $55M (FY11 $) |
| MO | 2012 | $60M; $30 M for balloons excluding launch (FY13 $) |
| SMEX | 2014 | $125M not including ELV, or $175M with PI-provided access to space (FY15 $) |
| MO | 2014 | $65M; $35M for suborbital class, including CubeSats (FY15 $) |
| MIDEX | 2016 | $250M, not including ELV (FY17 $) |
| MO | 2016 | $70M; $35M for suborbital class, including CubeSats (FY17 $) |

**Table S-2.** *Submitted Astrophysics Ex/MO proposals 2008 – 2016, shown with PI gender. This data is displayed in graphical form in **Figure 1** in the white paper* [1].

| Opportunity | Number of Proposals | | Total number of Proposals |
|---|---|---|---|
| | F PI | M PI | |
| 2008 MO | 1 | 10 | 11 |
| 2008 SMEX | 1 | 17 | 18 |
| 2011 EX | 0 | 15 | 15 |
| 2011 MO | 1 | 10 | 11 |
| 2012 MO | 1 | 8 | 9 |
| 2014 MO | 0 | 7 | 7 |
| 2014 SMEX | 0 | 13 | 13 |
| 2016 MIDEX | 0 | 9 | 9 |
| 2016 MO | 0 | 9 | 9 |
| **Grand Total** | **4** | **98** | **102** |

**Table S-3.** *Number and gender of PIs for submitted Astrophysics Ex/MO proposals 2008–2016*

|  | Total PIs | Unique PIs |
|---|---|---|
| Female | 4 | 3 |
| Male | 98 | 58 |
| **Total** | **102** | **61** |

**Table S-4.** *Science team size for Ex/MO proposals*

| Proposal Type | # Sci Team Members | | | |
|---|---|---|---|---|
|  | Min | Max | Median | Mean |
| MO | 6 | 42 | 15 | 17 |
| SMEX | 9 | 45 | 22 | 23 |
| MIDEX (includes EX) | 11 | 77 | 22 | 26 |

**Table S-5.** *Submitted and selected proposals in three Astrophysics ROSES elements, shown by PI gender. This data was compiled by D. Evans. Some of this data is displayed graphically in **Figure 7** in the white paper* [1].

| ROSES Element | Submitted Proposals | | | Selected Proposals | | |
|---|---|---|---|---|---|---|
|  | Total | % M PI | % F PI | Total | % M PI | % F PI |
| ADAP 2013–2015 | 832 | 78% | 22% | 163 | 79% | 21% |
| APRA 2012–2014 | 515 | 91% | 9% | 140 | 95% | 5% |
| ATP 2012–2014 | 586 | 82% | 18% | 84 | 87% | 13% |
| **Overall** | **1933** | **83%** | **17%** | **387** | **86%** | **14%** |

**Table S-6.** *Submitted and selected proposals for the initial RTF program, shown by PI gender. This table was composed by N. Barghouty.*

| Roman Technology Fellowship | Submitted Phase 1 Proposals | | | Selected Phase 1 Proposals | | | Selected Phase 2 Proposals | | |
|---|---|---|---|---|---|---|---|---|---|
| | Total # | # M PI | # F PI | Total # | # M PI | # F PI | Total # | # M PI | # F PI |
| 2011 | 19 | 17 | 2 | 3 | 3 | 0 | 2 | 2 | 0 |
| 2012 | 12 | 10 | 2 | 2 | 2 | 0 | 1 | 1 | 0 |
| 2013 | *RTF Program not offered* | | | | | | | | |
| 2014 | 8 | 7 | 1 | 3 | 3 | 0 | 2 | 2 | 0 |
| 2015 | 5 | 1 | 4 | 3 | 1 | 2 | 2 | 1 | 1 |
| **Total** | **44** | **35** | **9** | **11** | **9** | **2** | **7** | **6** | **1** |

**Table S-7.** *Submitted and selected proposals for the restructured RTF program, shown by PI gender. This table was composed by N. Barghouty.*

| Roman Technology Fellowship | APRA Submitted & RTF Qualified | | | APRA Selected & RTF Qualified | | | RTF Selected | | |
|---|---|---|---|---|---|---|---|---|---|
| | Total # | # M PI | # F PI | Total # | # M PI | # F PI | Total # | # M PI | # F PI |
| 2016 | 11 | 9 | 2 | 3 | 2 | 1 | 2 | 1 | 1 |
| 2017 | 7 | 5 | 2 | 3 | 3 | 0 | 3 | 3 | 0 |
| **Total** | **18** | **14** | **4** | **6** | **5** | **1** | **5** | **4** | **1** |

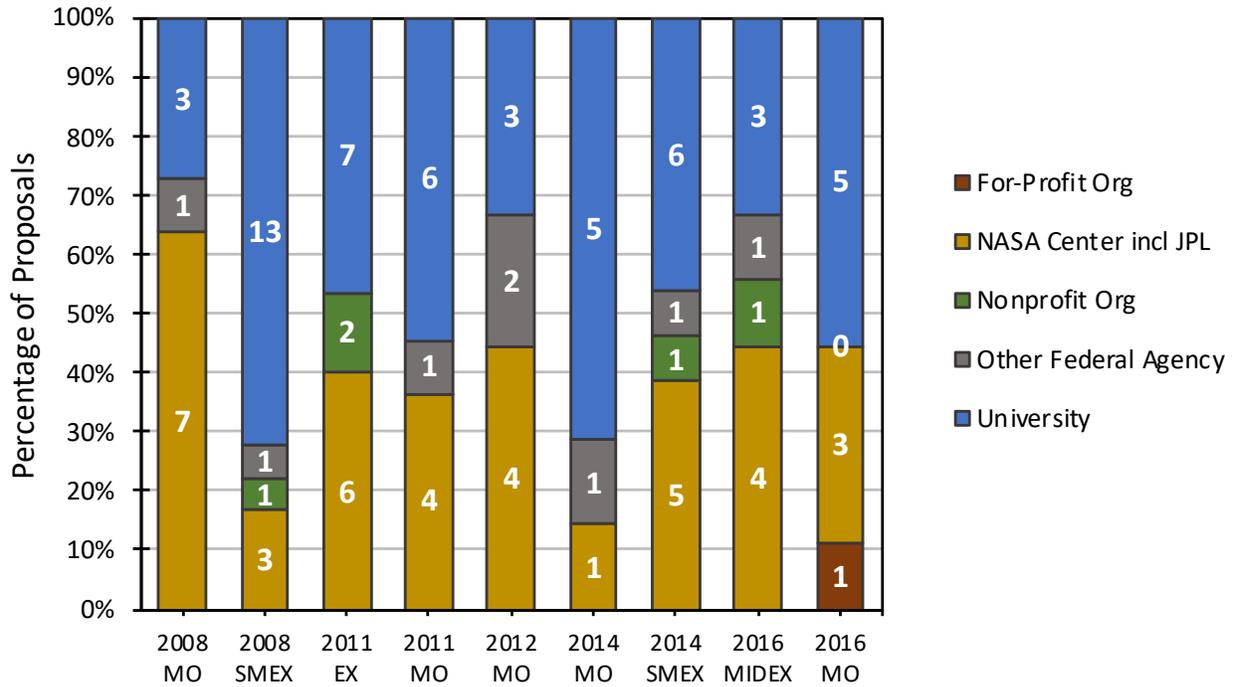

**Figure S-1.** *Number and percentage of Astro Ex/MO proposals 2008–2016 by opportunity and organization type. The column shown for each opportunity reaches to the 100% level to encompass all proposals submitted. Within each bar, the percentage of proposals submitted by each organization type is shown in the color-coded segments relative to the y-axis, and the number of proposals from each type is printed within the segments.*

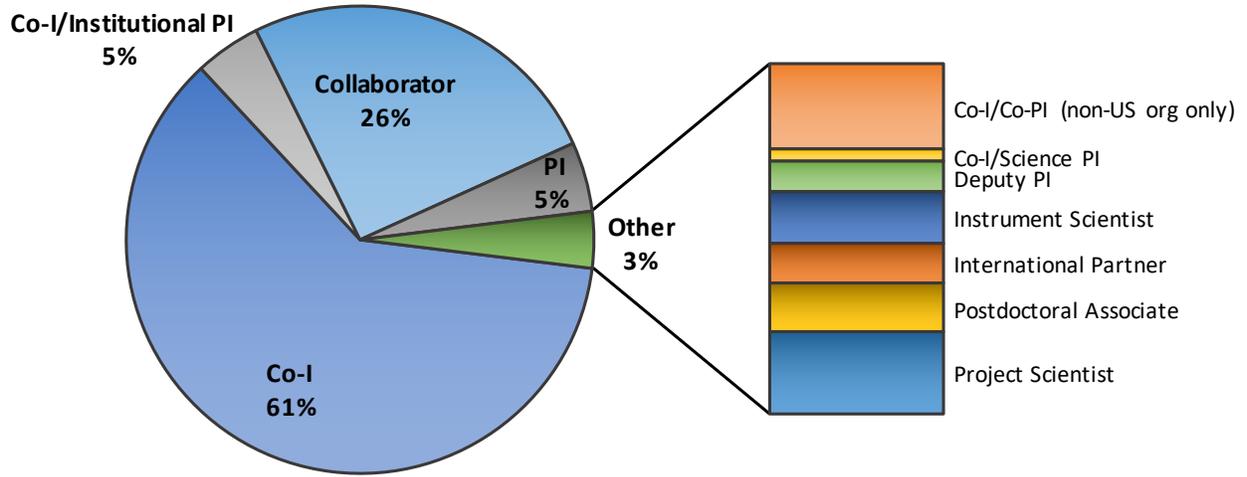

**Figure S-2.** *Total distribution of science roles (with both genders combined) for all Ex/MO proposals 2008–2016. This is a graphical representation of the data shown in the fourth column of **Table 1** in the white paper [1]. The total distribution of science roles among female participants and male participants taken separately (data in columns 2 and 3, respectively, in **Table 1** in [1]) do not show large scale differences from this plot.*

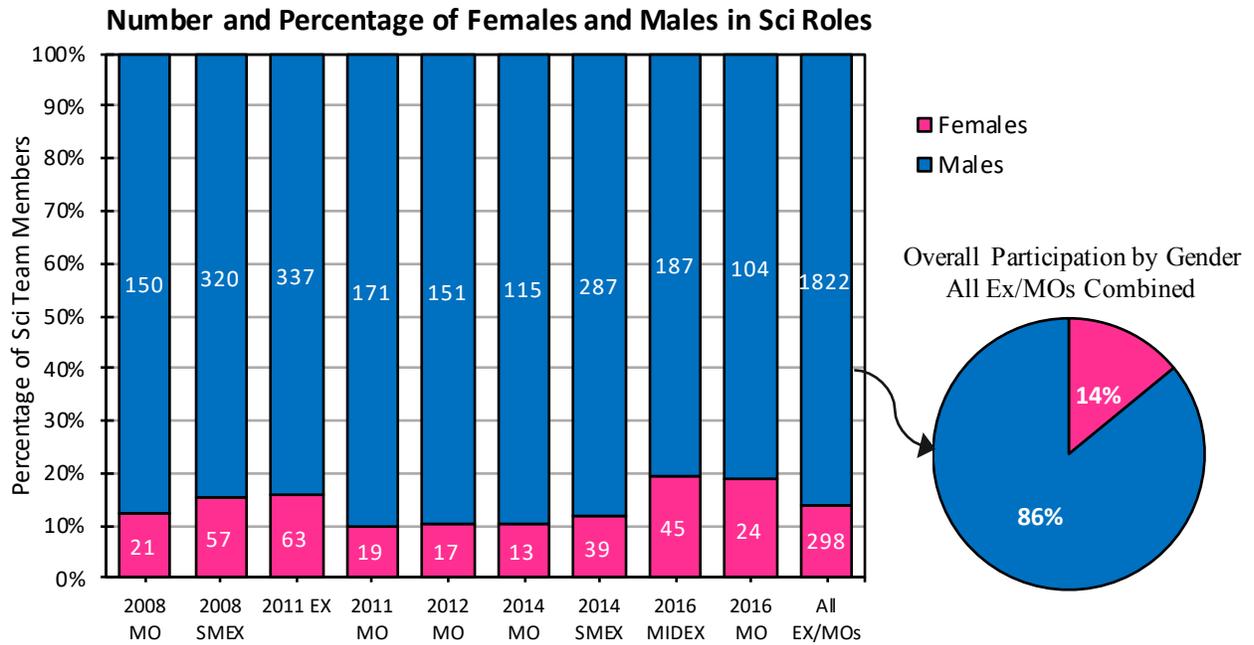

**Figure S-3.** *Overall number and percentage of participants in science roles is shown for Ex/MO opportunities. The column shown for each opportunity reaches to the 100% level to encompass all science participants. Within each bar, the percentage of participants by gender is shown in the color-coded segments relative to the y-axis, and the number of participants by gender is printed in each segment. The total participation by gender (rightmost column) is recapitulated in the pie chart.*

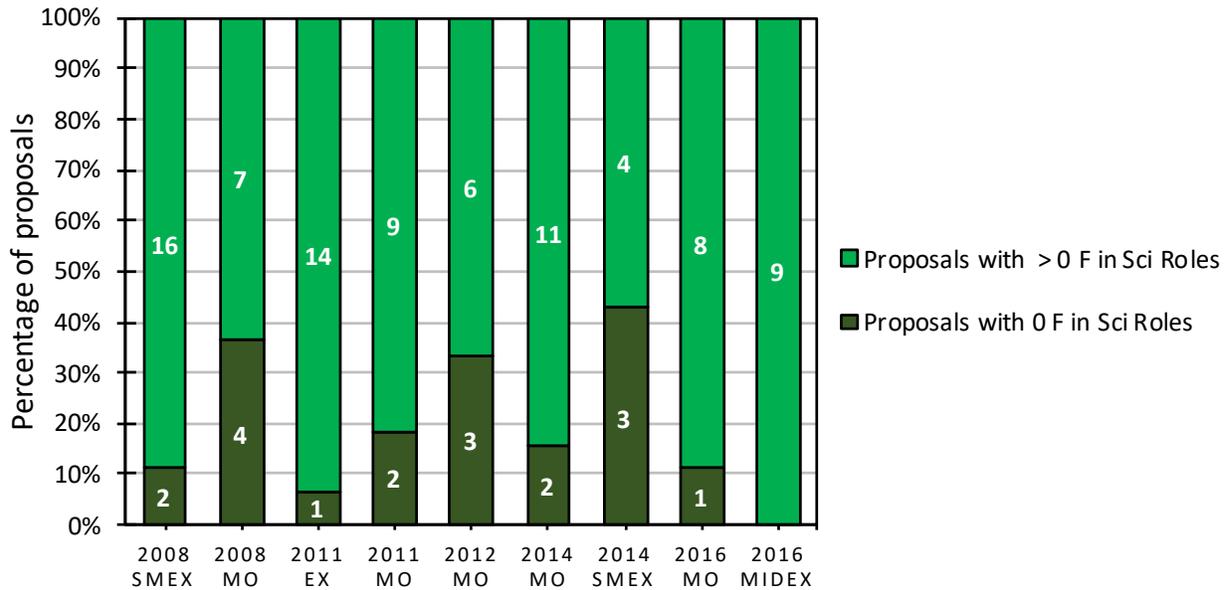

**Figure S-4.** *Number and percentage of Astro Ex/MO proposals having zero females in science roles submitted to AOs during 2008–2016. The column shown for each opportunity reaches to the 100% level to encompass all proposals submitted. Within each bar, the percentage of proposals submitted with either zero females or at least one female in a science role is shown in the color-coded segments relative to the y-axis, and the number of proposals in each of these cases is printed in the segments.*

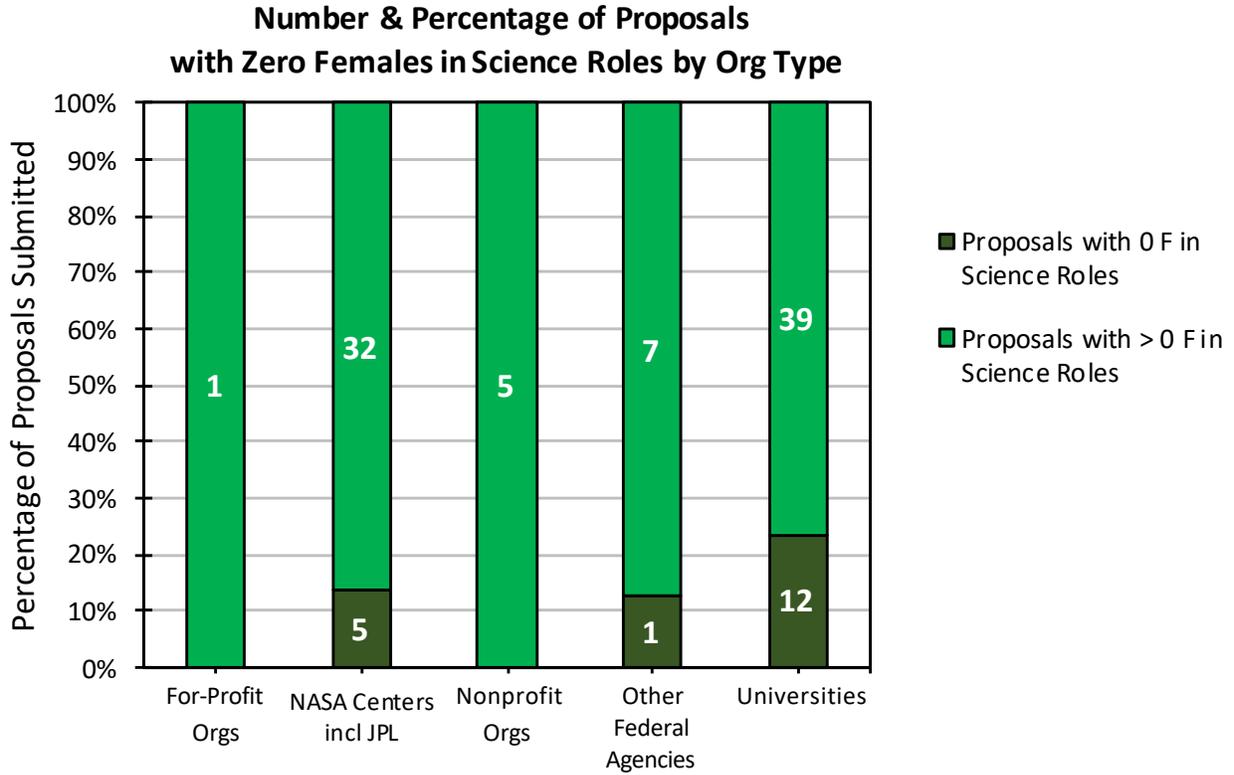

**Figure S-5.** *Number and percentage of Astro Ex/MO proposals having zero females in science roles. The column shown for each organization type reaches to the 100% level to encompass all proposals submitted. Within each bar, the percentage of proposals submitted with either zero females or at least one female in a science role is shown in the color-coded segments relative to the y-axis, and the number of proposals in each of these cases is printed in the segments.*

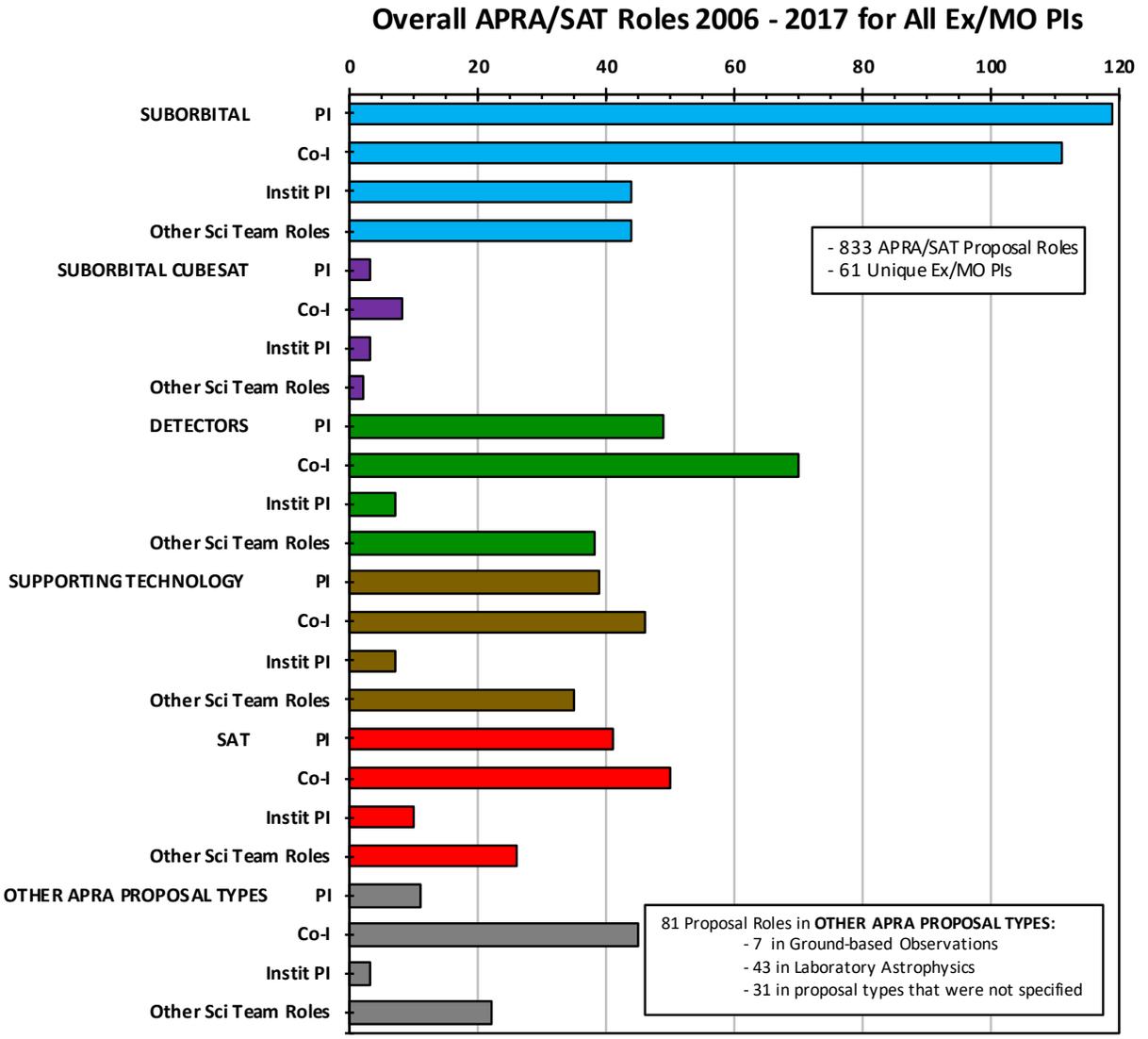

**Figure S-6.** *Bar chart showing all APRA/SAT roles held by the 61 unique Ex/MO PIs during 2006–2017.*